# Analysis of the mass structure of the hadrons

P. R. Silva – Retired Associated Professor - Departamento de Física- ICEx – Universidade Federal de Minas Gerais – Belo Horizonte – MG - Brazil
E-mail: prsilvafis@terra.com.br

**Abstract** – Inspired in previous works of Xiangdong Ji, published in PRL and PRD in 1995, we worked out an alternative way to separate within the structure of QCD, the hadron masses into contributions of quark and gluon kinetic and potential energies, quark masses and the trace anomaly. With respect to the nucleon mass the present results are between the two approximations developed by Ji. We also developed three approximations for the separation of the pion mass, which is also compared with Ji results. With the help of the quark condensate relation obtained by Nassif and Silva in 2006, we were able to separate the quark energy into its kinetic and potential parts.

## 1 – INTRODUCTION

In a letter dealing with the mass structure of nucleon, Ji [1] states that an insight on this subject can be achieved by using QCD with help of the deep-inelastic momentum sum rule and the trace anomaly.

As was pointed out by Ji [2] for any physical system, a good knowledge about its mass structure is helpful in understanding the underlying dynamics of it. Through a study of the energy-momentum tensor of QCD, Ji [1,2] was able to make the separation of the hadron masses into contributions from quark kinetic and potential energies, the gluon energy, the current quark masses, and the trace anomaly. Also according to Ji [2], the part of the trace anomaly is a direct consequence of the scale symmetry breaking, and is analogous to the vacuum pressure empirically introduced in the MIT bag model [3,4]. In order to evaluate the separation of the masses terms, Ji [2] makes use of two matrix elements: the momentum fraction of the hadron carried by quarks in the finite momentum frame and the quarks scalar charges.

In the present work we intend to pursue further on this subject, namely to estimate the mass separation in hadrons, through alternative paths to those followed by Ji [1,2]. A way of doing is by making the direct evaluations of the contributions, for instance by estimating the strong couplings of the nucleon and the pion, just at their respective rest masses. Besides this, by assuming reasonable ratios between the color-electric and color-magnetic field intensities, it is possible to make direct evaluations of the gluon energy and of the trace anomaly contributions to the hadron masses.

As an phenomenology oriented author, we skip the detailed calculations worked out by Ji [1,2] which shows that the energy-momentum tensor of QCD can be decomposed between traceless and trace parts. According to Ji [1,2] the traceless part of it can be separated into contributions of the quark and gluon parts. On the other hand, the trace part of the energy-momentum tensor of QCD is decomposed into the quark mass and the trace anomaly terms. We refer to the Physical Review Letters [1] and Physical Review D [2] papers by Xiangdong Ji, for detailed analysis of these separations.

## 2 – SOME PRELIMINAIRES

According to Xiangdong Ji [1], the QCD Hamiltonian reflecting the mass separation of the hadrons can be written as

$$H_{QCD} = H_q + H_m + H_g + H_a, \tag{1}$$

where

$$H_q = \int d^3\mathbf{x}\ \Psi^+(-i\mathbf{D}.\boldsymbol{\alpha})\Psi, \tag{2}$$

$$H_m = \int d^3\mathbf{x}\ \Psi^+\ m\ \Psi, \tag{3}$$

$$H_g = \int d^3\mathbf{x}\ \tfrac{1}{2}(\mathbf{E}^2 + \mathbf{B}^2), \tag{4}$$

$$H_a = \int d^3\mathbf{x}\ [9\alpha_s/(16\pi)]\ (\mathbf{E}^2 - \mathbf{B}^2). \tag{5}$$

Here $H_q$ [eq. (2)] stands for the quark and the anti-quark kinetic and potential energies (quark energy for short) and contributes with $3(a-b)/4$ fraction of the hadron mass M. $H_m$ [eq.(3)] is the hadron mass term and contributes with b fraction of the mass. $H_g$ [eq.(4)] is the gluon energy and contributes with $3(1-a)/4$ fraction of the hadron mass. Finally, $H_a$ [eq.(5)] is the trace anomaly term and contributes with $(1-b)/4$ fraction of the hadron mass.

As was pointed out by Ji [1, 2], in the chiral limit, the gluon energy of the trace anomaly (equal to $M/4$) corresponds exactly to the vacuum energy in the MIT bag model [3,4]. Therefore the chiral limit is equivalent to take $b = 0$ and $a = .5$ in the previous representations of the fractions of the hadron mass. X Ji [1,2] evaluated the color-electric and color-magnetic fields contributions for the gluon energy and found that the magnetic-field energy is negative. It seems that this negative energy could be attributed to the gluon color paramagnetism.

3 – THE MASS STRUCTURE OF THE NUCLEON

Inspired in X Ji [1,2] work, taking into account the negative contribution of the color-magnetic field to the gluon energy, and starting from the chiral limit we write

$$M_g/M_a = \{\tfrac{1}{2}(\mathbf{E}^2 - [\mathbf{B}^2])\ V\}/\{[9\alpha_s/(16\pi)]\ (\mathbf{E}^2 + [\mathbf{B}^2])\ V\} = 3/2. \tag{6}$$

In eq.(6), $\mathbf{E}^2$ stands for the averaged color-electric field and V is the volume of the nucleon. We also took the ratio between the gluon energy and the trace anomaly energy as $3/2$, by considering the chiral limit. Besides this we used the algebraic trick

$$[\mathbf{B}^2] = -\mathbf{B}^2. \tag{7}$$

The algebraic trick used in (7) is a way of dealing with the negative color-magnetic field contribution for the gluon energy.

In order to proceed further we are going to determine the strong coupling strength $\alpha_s$ at the energy scale of the nucleon. Let us take the color-charge of the nucleon homogeneously distributed over a spherical shell of radius r plus a surface tension term σ, in an analogous way to that considered in the Dirac's extensible model of the electron [5, 6], leading to the potential

$$P_{Dirac} = \alpha_s \hbar c/(2r) + 4\pi r^2 \sigma. \tag{8}$$

The radius of equilibrium of this spherical shell is given by

$$R_s = (3\alpha_s \hbar)/(4Mc). \qquad (9)$$

Naturally, due to the Gauss' law, the color-electric field is null inside the shell and finite outside it. This behavior of the color-electric fields mimics a step-like function behavior of the running coupling constant, namely

$$\alpha_s(r \geq R_s) = \text{finite}, \qquad \text{and} \quad \alpha_s(r \leq R_s) = 0. \qquad (10)$$

The null intensity of the color-electric field inside the shell resembles the asymptotic freedom behavior of the running coupling constant (Please see reference [7]). Now we are led to think this spherical shell of radius $R_s$ with the consequent discontinuity it imposes on the color-electric field, behaving as an event horizon surface. In this way it is possible to attribute to it a kind of Hawking radiation. In the present deduction we follow the steps undergone in reference [8], dealing with the physics of the black hole radiation.

Let us consider at the surface of the spherical shell (resembling a surface horizon of a black hole), a circle of radius $R_s$ (eq.(9)). A fluctuating color-electric field that can radiate will be taken as a transverse field of intensity $E_t$, and we suppose it has equal probability to assume positive or negative values. A color test charge $e_s$ coupled to this field and the integration of this force over a small interval of the circle's arc $\Delta S$ will give on average null contribution for the energy, due to the fluctuating character of this field. Thus we need to take into account the second moment of this "elastic" energy and we write

$$U = [e_s E_t(\Delta S/2\pi)]^2/(2Mc^2) \qquad (11)$$

In (11), we have "normalized" U dividing the numerator by $2Mc^2$, by taking into account the possibility of producing a particles pair from the vacuum. The field $E_t$ is assumed to be responsible by the nucleon excitation. We want to compare this field with the radial color-electric field $E_r$, which maintains the test charge tied to the spherical shell. We also rescaled $\Delta S$, dividing it by $2\pi$ (eq.(11)). In this way we assure that a light front will take the same time of travel to go from the center of symmetry of the shell to its surface, as to cover its rescaled perimeter. We can also write

$$U = \tfrac{1}{2} k(\Delta S)^2, \qquad (12)$$

where

$$k = (e_s E_t)^2/[(2\pi)^2 Mc^2], \qquad (13)$$

is the spring constant of a harmonic oscillator. The angular frequency of this oscillator is

$$\omega = (k/M)^{1/2} = (e_s E_t)/(2\pi Mc). \qquad (14)$$

Solving (14) for $E_t$ we get

$$E_t = (2\pi\omega Mc)/e_s. \tag{15}$$

On the other hand, the radial field at the spherical surface is given by

$$E_r|_{r=R_s} = (1/4\pi\varepsilon_o) e_s/R_s^2 = (1/4\pi\varepsilon_o) e_s M^2 c^2/[(9/16)\alpha_s^2 \hbar^2]. \tag{16}$$

In (16) we wrote the strong (or color) charge $e_s$ in terms of an equivalent electric charge. We assume that at the surface of the spherical shell, considered as a surface horizon, the vacuum fluctuations are so great that the strength of the radiative field equals to that of the radial binding field. So, by imposing the equality between $E_t$ and $E_r$ fields (eqs.(15) and (16)), we get

$$\hbar\omega = (8Mc^2)/(9\pi\alpha_s). \tag{17}$$

To obtain (17) we also used that

$$e_s^2/(4\pi\varepsilon_o) = \alpha_s \hbar c. \tag{18}$$

Alternatively, the use of the Bohr-Sommerfeld method of quantization on treating the radial excitations of the nucleon, led to the separation in its energy levels, namely the separation between the ground state and the centroid of the excited states [9, 10]. The obtained result [9] was

$$\Delta U = (2Mc^2)/\pi. \tag{19}$$

By imposing the equality between $\hbar\omega$ (eq. (17)) and $\Delta U$, we finally obtain

$$\alpha_s = 4/9. \tag{20}$$

This is the value we have estimated for the strong coupling at the energy scale of the nucleon.
Inserting the value of $\alpha_s$ given by (20) into (6) and solving for the ratio $[\mathbf{B}^2]/\mathbf{E}^2$, in the chiral limit, we get

$$[\mathbf{B}^2]/\mathbf{E}^2 = (4\pi - 3)/(4\pi - 3) = .614 \approx 6/10. \tag{21}$$

We must to stress that (21) stands for the ratio between the averaged magnitudes of the magnetic-color and electric-color fields.
Pursuing further within the chiral limit we can write

$$(1/2)(\mathbf{E}^2 - [\mathbf{B}^2])V = (3/8)M, \tag{22}$$

and by using (21) we obtain

$$\langle P|\mathbf{E}^2|P\rangle = \mathbf{E}^2 V = 1826 \text{ MeV}, \tag{23}$$

and

$$\langle P|\mathbf{B}^2|P\rangle = \mathbf{B}^2 V = -1121 \text{ MeV}. \tag{24}$$

The above results must be compared with those evaluated by X Ji [1, 2] which got 1700 MeV and – 1050 MeV for the color-electric and color-magnetic fields respective contributions to the gluon energy part of the nucleon mass.

Next we intend to improve the previous calculations of the mass structure of the nucleon, starting from relation (21) and going beyond the chiral limit.

## 4 – THE MASS STRUCTURE OF THE NUCLEON BEYOND THE CHIRAL LIMIT

Taking into account eq.(21), we can write

$$\int d^3\mathbf{x}\, \tfrac{1}{2}(\mathbf{E}^2 + \mathbf{B}^2) \approx (1/2)(.4)\{\mathbf{E}^2(r,t)\}|_{\text{space-time average}} V. \qquad (25)$$

The next step it is to evaluate the averages in the color-electric field. To accomplish this, let us make the following reasoning. A known feature of the strong force behavior is that it has a short-range order character, represented by the color fields being confined in the region inside the nucleon volume, as in the MIT bag model [3,4] case. A possible mimic to represent this behavior is to take the strong charge of the nucleon homogeneously distributed over a spherical volume of radius R, and with an effective strong coupling given by $\alpha_{\text{eff}}$. Besides this we need a spherical shell with an equal amount of "anti-strong charge" also of radius R, so that Gauss' law implies in the exact cancellation of the effective color-electric field in the region r > R. A straightforward calculation of the energy stored in this sphere yields

$$U_{\text{static}} = (1/10)\, (\alpha_{\text{eff}}\, \hbar c)/R. \qquad (26)$$

Turning to the evaluation of the averages of the electric-color field we have

$$\{\mathbf{E}^2(r,t)\}|_{\text{space-time}} = \{\mathbf{E}^2(r)\}|_{\text{space}} <\cos^2\omega t>|_{\text{time}} = (1/2)\{\mathbf{E}^2(r)\}|_{\text{space}}. \qquad (27)$$

In eq.(27), we have written the time dependence of the electric-color field in terms of one mode of oscillation, namely $\cos^2\omega t$, leading to a $1/2$ factor as a consequence of the time-averaging. For the space dependence we can write

$$\{\mathbf{E}^2(r,t)\}|_{\text{space-av}} = E_0^2 (1/R^3)\int_o^R r^2\, dr = E_0^2/3. \qquad (28)$$

In eq.(28), we took into account the behavior of a static electric field of a homogeneous sphere of charge. Thus using equations (28), (27) and (25), we get

$$M_g = (1/3)(.1\, E_0^2 V) = (1/3)\, M, \qquad (29)$$

where we have identified $(E_0^2 V)/10$, with the nucleon mass M, after looking at (26). Beyond the chiral approximation (BCA), the trace anomaly contribution for the nucleon mass reads

$$M_a = \int d^3\mathbf{x}\, [9\alpha_s/(16\pi)]\, (\mathbf{E}^2 + [\mathbf{B}^2]) = 4(1.6)/(16\pi)\{\mathbf{E}^2(r,t)\}|_{\text{space-time}} V. \qquad (30)$$

Working in an analogous way we have done before we obtain

$$M_a = (2/3\pi)(.1\, E_0^2 V) = (2/3\pi)\, M. \tag{31}$$

We observe that for sake of simplicity, we took the ratio between the absolute values contributions of the electric and magnetic color fields as exactly equal to $6/10$. Now by using X Ji [1,2] notation, we can write

$$M_a = (1-b)\, M/4 = (2/3\pi)\, M, \tag{32}$$

which gives

$$M_m = bM = [(3\pi - 8)/(3\pi)]M. \tag{33}$$

We also have

$$M_g = 3(1-a)M/4 = M/3, \tag{34}$$

leading to

$$a = 5/9 \quad \text{and} \quad b = (3\pi - 8)/(3\pi) \approx .151. \tag{35}$$

Then the quark kinetic plus potential energy is given by
$$M_q = 3(a - b)M/4 = [(6 - \pi)/(3\pi)]M. \tag{36}$$

We present in table 1, the separation of the nucleon mass into different contributions evaluated in the BCA. There, our calculations are compared with those evaluated by X Ji [1,2].

Table 1 – Nucleon mass separated into different contributions. This work [beyond chiral approximation (BCA)] is compared with X. Ji [1,2] results.

| Mass Type | $M_i$ | $m_s \to 0$ (MeV) | $m_s \to \infty$ (MeV) | Present work (MeV) |
|---|---|---|---|---|
| Quark energy | $3(a-b)/4$ | 270 | 300 | 285 |
| Quark mass | $b$ | 160 | 110 | 142 |
| Gluon energy | $3(1-a)/4$ | 320 | 320 | 313 |
| Trace Anomaly | $(1-b)/4$ | 190 | 210 | 199 |

Looking at table 1, we notice that the results of this work (BCA) are between those evaluated by considering extreme values of the strange quark mass. May be they indirectly reflect the finiteness of the strange quark mass.

By considering the gluon energy we can write

$$\tfrac{1}{2}(.4)\{\mathbf{E}^2\}|_{\text{space-time av.}} V = M/3, \tag{37}$$

where we have taking into account (29) and that $[\mathbf{B}^2] = .6\, \mathbf{E}^2$. Therefore we get

$$<P|\ \mathbf{E}^2\ |P>|_{BCA} = \{\mathbf{E}^2\}|_{\text{space-time av.}}\ V = (5/3)\ M = 1565\ \text{MeV}, \tag{38}$$

and

$$<P|\ \mathbf{B}^2\ |P>|_{BCA} = \{\mathbf{B}^2\}|_{\text{space-time av.}}\ V = -\ M = -\ 939\ \text{MeV}. \tag{39}$$

BCA stands for Beyond Chiral Approximation.

Taking the sum of the color-electric and the color-magnetic contributions to the energy we obtain

$$<P|\ \mathbf{E}^2\ |P>|_{BCA} + <P|\ \mathbf{B}^2\ |P>|_{BCA} = (2/3)\ M = \hbar\omega_{HOA}. \tag{40}$$

In (40), $\omega_{HOA}$ stands for the frequency of the harmonic oscillator approximation, sometimes used to describe the nucleon mass [10].

5 – THE MASS STRUCTURE OF THE PION

In order to evaluate the contributions to the mass of the pion, let us start from the MIT bag model [3,4,9]. It is possible to think about a characteristic sound velocity $v_s$, related to a massive particle of color-charge $e_s$. We write

$$v_s = (P_{vac}/\rho)^{1/2} = (½)\ c, \tag{41}$$

where $P_{vac}$ and $\rho$ are the hadron pressure and the averaged density of hadron matter. On the other hand let us consider the Lorentz color-force acting on this particle. We have

$$d\mathbf{p}/dt = e_s\ \mathbf{E} + (e_s/c)\ \mathbf{v_s}\ \mathbf{x}\ \mathbf{B}, \tag{42}$$

where $\mathbf{p}$ is the momentum of the dressed quark or dressed gluon. In a representation where color-electric and color-magnetic fields have the same dimensionality, the energy associated to electric part is proportional to $e_s^2$ and to the magnetic one is proportional to $(e_s v_s/c)^2$. Also due to the color- paramagnetism of the vacuum the magnetic contribution has a negative signal. Basing on these ideas we can write

$$<\mathbf{B}^2>|_{\text{time av.}} = -\ ¼\ <\mathbf{E}^2>|_{\text{time av.}} \tag{43}$$

As a means to deal with the separation of the pion mass it seems better to assume the pion modeled as a flux tube, rather than it endowed with a spherical symmetry. Therefore we write for the gluon energy contribution to the pion mass:

$$M_g = ½\ (\ <\mathbf{E}^2>|_{\text{time av.}} +\ <\mathbf{B}^2>|_{\text{time av.}})\ V = (½)(3/4)\ <\mathbf{E}^2>|_{\text{time av.}}\ V. \tag{44}$$

But

$$<\mathbf{E}^2>|_{\text{time av.}} = E_0^2\ <\cos^2\omega t>|_{\text{time av.}} = ½\ E_0^2, \tag{45}$$

yielding

$$M_g(\text{pion}) = (3/8)\ (½E_0^2 V) = (3/8)\ m_\pi. \tag{46}$$

In (47), we have identified $\frac{1}{2}E_0^2 V$ with $m_\pi$.

In the next step we will evaluate the trace anomaly contribution for the pion mass. This will be done through two approximations.

TRACE ANOMALY OF THE PION - FIRST APPROXIMATION

To evaluate the trace anomaly contribution to the pion mass we need to know the strong coupling constant at this energy scale. One way to do this is by observing a plot of the Cornell Potential, where we verify that there is a very good agreement among the various numerical fittings used to represent it (please see A. V. Nesterenko paper [11]. We also observe that the zero of these fittings to the Cornell Potential (CP) occurs at approximately .5 fm. Meanwhile the "Maximum Floatability Hypothesis" was applied to the CP (please see Silva [12]) and was found that the zero of it occurs at the radius $R_0$ given by

$$R_0 = [(4/3)\alpha_s \hbar]/(m_q c), \qquad (47)$$

where $m_q$ is the quark constituent mass, taken as one third of the nucleon mass. By taking $R_0 = .5$ fm, and solving (47) for $\alpha_s$ we find that

$$\alpha_s \text{ (pion)} = .591. \qquad (48)$$

The energy contribution of the trace anomaly for the pion mass can therefore be evaluated in this first approximation. We have

$$M_a = [(9\,\alpha_s)/(16\pi)](<P|\mathbf{E}^2 - \mathbf{B}^2|P>) = [(9\,\alpha_s)/(16\pi)](<\mathbf{E}^2> - <\mathbf{B}^2>)|_{\text{time av.}}. \qquad (49)$$

Using (43), (45) and (48) into (49), we finally obtain

$$M_a \text{ (pion)} \approx (1/7.6)\, m_\pi. \qquad (50)$$

Relation (50) implies that, in a first approximation, we have

$$b \approx 9/19, \qquad (51)$$

and as we get from (46)

$$a = 1/2. \qquad (52)$$

These evaluated results are used to glimpse an anatomy of the pion mass in Table 2. There we are going to compare this first approximation for the separation of the pion mass with the results obtained by Ji [2]. Now let us go to the second approximation.

TRACE ANOMALY OF PION – SECOND APPROXIMATION

As was pointed out by X Ji [2], according to the Goldstone theorem, the pion is intrinsically different from ordinary hadrons: - it is a collective mode in QCD vacuum.

Based on this statement we are going to assume that the quarks kinetic plus potential energy and quarks masses both contribute with zero for the pion mass. To accomplish this, we will set a = b = 0 in the X Ji relations [1,2], in order to determine the strong coupling at the pion mass scale, in the second approximation. We write

$$\tfrac{1}{2} ( <\mathbf{E}^2>|_{\text{time av.}} + <\mathbf{B}^2>|_{\text{time av.}}) \, V = (3/4) \, m_\pi, \qquad (53)$$

and

$$[(9 \, \alpha_s)/(16\pi)](<\mathbf{E}^2> - <\mathbf{B}^2>)|_{\text{time av.}} \, V = (1/4) \, m_\pi. \qquad (54)$$

By using (43) in the (53) and (54) relations and solving for $\alpha_s$, we find

$$\alpha_s(\text{pion})|_{\text{second approx.}} = (8\pi)/45 \approx .559. \qquad (55)$$

Now, by using (43), (45) and (55) into (49) we obtain

$$M_a \, (\text{pion})|_{\text{second approx.}} = (1/8) \, m_\pi. \qquad (56)$$

As can be verified in Table 2, we got in this second approximation the same numbers for the mass structure of pion as those obtained by X Ji [2]. Nevertheless, we can also determine separately the color-electric and the color-magnetic field contributions and we get

$$<P| \, \mathbf{E}^2 \, |P> = \{\mathbf{E}^2\}|_{\text{space-time av.}} \, V = m_\pi, \qquad (57a)$$

$$<P| \, \mathbf{B}^2 \, |P> = \{\mathbf{B}^2\}|_{\text{space-time av.}} \, V = - (\tfrac{1}{4}) m_\pi. \qquad (57b)$$

These estimations, the separated contributions of color-electric and color-magnetic fields to the gluon energy, differs from those obtained by X Ji [2], where for simplicity, pure non-Abelian gauge theory (without quarks) was considered.

TRACE ANOMALY OF PION – THIRD APPROXIMATION

Let us start from the Goldberger-Treiman relation [13]

$$g_{\pi qq} \, f_\pi = m_q, \qquad (58)$$

where $g_{\pi qq}$ is the pion-quark-quark coupling and $f_\pi$ is the pion constant. Besides this we can also write

$$g_{\pi qq} \, m_\pi = M/2 = (3/2) \, m_q. \qquad (59)$$

In (59) we supposed the nucleon interacting with itself through a mix of quarks and the pion. Combining (58) and (59) yields

$$m_\pi = (3/2) \, f_\pi. \qquad (60)$$

Now let us consider the possibility that the, $<\mathbf{B}^2>|_{\text{time av.}} = -\tfrac{1}{4}<\mathbf{E}^2>|_{\text{time av.}}$ relation, namely eq. (43), used in the first and second approximations to evaluate the mass structure of pion could be modified, perhaps even in a certain sense improved. Then we write

$$<\mathbf{E}^2> V = m_\pi, \tag{61a}$$

$$(<\mathbf{B}^2> + <\mathbf{E}^2>) V = f_\pi = (2/3) m_\pi. \tag{61b}$$

Relations (61) imply that

$$<\mathbf{B}^2> V = -(1/3) m_\pi = -(1/3) <\mathbf{E}^2> V. \tag{62}$$

Using (62) and $\alpha_s(\text{pion})$ given by (55), we can evaluate again the gluon energy and trace anomaly parts of the pion mass, obtaining

$$M_g(\text{pion})|_{\text{third app.}} = (1/3) m_\pi, \tag{63a}$$

$$M_a(\text{pion})|_{\text{third app.}} = (2/15) m_\pi. \tag{63b}$$

The third approximation leads to

$$a = 5/9, \quad \text{and} \quad b = 7/15. \tag{64}$$

In table 2, the three approximations developed in this work for analyze the mass structure of the pion are compared with that worked out by X Ji [2]. We notice that there is an agreement between the second approximation of this paper and that of X Ji work.

Table 2 – Pion mass separated into different contributions. Three approximations developed in this work are compared with X. Ji [2] results.

| Mass Type | $M_i/m_\pi$ | This work First Appr. | This work Second Appr. | This work Third Appr. | X Ji work Reference [2] |
|---|---|---|---|---|---|
| Quark energy | $3(a-b)/4$ | 3/152 | 0 | 1/15 | 0 |
| Quark mass | b | 9/19 | 1/2 | 7/15 | 1/2 |
| Gluon energy | $3(1-a)/4$ | 3/8 | 3/8 | 5/15 | 3/8 |
| Trace Anomaly | $(1-b)/4$ | 1/7.6 | 1/8 | 2/15 | 1/8 |

6 – DIRECT EVALUATION OF THE QUARK MASS

It is possible to directly evaluate the quark mass by making the following assumptions. First we write

$$\int d^3\mathbf{x}\, \Psi^+ m\, \Psi = |<\Psi^+\Psi>|\, m\, V, \tag{65}$$

where

$$|<\Psi^+\Psi>| = (\pi\, M^3)/128, \tag{66}$$

Is the absolute value of the quark condensate as obtained in reference [7]. WE put by the hand the bare mass sum contribution to the nucleon

$$M \approx 20\ \text{MeV} \approx M/47, \tag{67}$$

and the nucleons volume V equal to

$$V = (4/3)\, \pi\, R^3 = (4/3)\, \pi\, (4/M)^3. \tag{68}$$

In (68) we took the nucleon radius equal to $4/M$, as estimated by X Ji [14].
  Inserting these data on the quark mass contribution to the nucleon, we get

$$M_q = (2\pi^2)\, M/141, \tag{69}$$

yielding

$$b = (2\pi^2)/141 \approx .140. \tag{70}$$

7 – BREAKING OF THE QUARK ENERGY: KINETIC PLUS POTENTIAL

  In this section we will show that it is possible to evaluate separately the quark kinetic and potential energies contributions for the nucleon mass. Starting from the relation

$$H_q = \int d^3\mathbf{x}\, \Psi^+(-i\mathbf{D}\cdot\boldsymbol{\alpha})\Psi, \tag{71}$$

and by considering

$$D^\mu = \partial^\mu + i\, g\, A^\mu, \tag{72}$$

we can write

$$H_q = \int d^3\mathbf{x}\, \Psi^+(-i\, \partial/\partial r - gA)\Psi, \tag{73}$$

where some simplification was made taking into account that a kind of averaging will be done at the end of the calculations. Besides this we make the approximation

$$\Psi = \Psi_0\, e^{i\, r/R}. \tag{74}$$

Therefore we have for the quark kinetic energy part of the nucleon mass the relations

$$<K> = \int d^3\mathbf{x}\, \Psi^+(1/R)\Psi = |<\Psi^+\Psi>|\, (1/R)\, V = (\pi^2/6)\, M. \tag{75}$$

In the above development, we have considered the relation for the quark condensate as given in [7] and the volume V of a sphere with the nucleon's radius, $R = 4/M$ [3].

THE POTENTIAL ENERGY

Within the Harmonic Oscillator Approximation (HOA), we can suppose that the absolute value of the potential energy is responsible by half of the quark "interaction mass". Hence we write ($\hbar = c = 1$)

$$m_{int}|_{HOA} = -(2\alpha_s)/R. \qquad (76)$$

Beyond the HOA, we can write

$$m_{int}|_{BHOA} = -(2f\alpha_s)/R, \qquad (77)$$

where f is a number to be determined. In the pion case, by considering the second approximation, we can take into account the bare contributions to the kinetic and potential energies, namely

$$1/R[1 - 2f\alpha_s(\text{pion})] = 0. \qquad (78)$$

But in the second approximation $\alpha_s(\text{pion}) = (8\pi/45)$, which implies

$$2f = 45/(8\pi). \qquad (79)$$

On the other hand, in the nucleon case, we have

$$-g A = m_{int}|_{BHOA} = -[2f\alpha_s(\text{nucleon})]/R. \qquad (80)$$

By using $\alpha_s(\text{nucleon}) = 4/9$ and $2f = 45/(8\pi)$, we obtain

$$-g A = m_{int}|_{BHOA} = -(1/R) 15/(6\pi). \qquad (81)$$

Therefore the quark potential energy contribution to the nucleon mass is

$$<P_{ot}> = -g A | <\Psi^+\Psi> | V = -[(15\pi)/36] M. \qquad (82)$$

We may test this result by putting

$$M_q = <K> + <P_{ot}> = [(\pi^2/6) - (15\pi)/36] M = (3/4)(a - b) M. \qquad (83)$$

Thus by taking $a = 5/9$, we obtain

$$b = (5 + 5\pi - 2\pi^2)/9 \approx .11. \qquad (84)$$

8 - THE BARE MASS OF THE QUARKS IN THE NUCLEON

It is possible to use the value of b estimated in (84) in order to evaluate the bare mass of the quarks in the nucleon. Starting from

$$H_m = \int d^3\mathbf{x}\, \Psi^+ m \Psi = b\, M, \qquad (85)$$

and taking

$$m = M/\gamma, \qquad (86)$$

and b given by (84), we obtain

$$\gamma = 6\pi^2/(5 + 5\pi - 2\pi^2) \approx 61. \qquad (87)$$

The above result leads to

$$m = M/\gamma \approx 15.4 \text{ MeV}. \qquad (88)$$

The result (88) for the bare mass of quarks in the nucleon must be compared with the value of 20 MeV considered before, which was inserted in the $H_m$ by the hand.

9 – RADIUS OF THE PION FROM THE QUARK MASS

A way to evaluate the pion radius could be through these steps. First we write the relation for the quark mass contribution to the pion mas

$$M_m(\text{pion}) = \int d^3\mathbf{x}\, \Psi^+ m \Psi = m \langle \Psi^+ \Psi \rangle V_\pi. \qquad (89)$$

Now we use the harmonic oscillator approximation to represent he pion mass

$$m_\pi = (3/2) f_\pi = (3/2)\, \omega_{HOA}, \qquad (90)$$

where we have identified $f_\pi$ with the frequency of the harmonic oscillator. To pursue further we consider the Gell-Mann, Oakes, Renner relation [15]

$$f_\pi^2 m_\pi^2 = m\, |\langle \Psi^+ \Psi \rangle|. \qquad (91)$$

We also write

$$R_\pi = \delta/m_\pi. \qquad (92)$$

and

$$M_m(\text{pion}) = b\, m_\pi = (1/2)\, m_\pi. \qquad (93)$$

Working with the above relations we find that

$$\delta = 3/(32\pi)^{1/3} \approx 3/4.65. \qquad (94)$$

Finally we obtain for the pion radius

$$R_\pi = \delta/m_\pi \approx .91 \text{ fm}. \tag{95}$$

## 10- ANALOGY WITH THE CRYSTAL GROWING

Consider the following relation

$$H_m(\text{pion}) = \int d^3\mathbf{x}\, \Psi^+ m\, \Psi = m|\langle\Psi^+ \Psi\rangle| V_\pi. \tag{96}$$

It is tempting to make an analogy between the above relation and the growing of a crystal from a water solution of its salt. If we consider that we have the bare mass, a sum of a two-flavor quark-antiquark pair, approximately equal to $m_\pi/14$ or 10 MeV, we obtain as an output the mass contribution of the quarks to the pion equal to $m_\pi/2$ or 70 MeV.

In the process of crystal growing, starting from a saturated water solution of the salt, a small seed is merged into the solution and slowly the crystal proceeds to grow by adding mass of the solution to the seed.

We may think that the vacuum from the quantum chromodynamics, represented by the quark condensate, works in a similar way to the water solution of the salt, and dresses the bare mass of the pion, multiplying its value by a factor of seven. This analogy can also be applied to the nucleon case.